\documentclass[runningheads]{llncs}
\usepackage[T1]{fontenc}
\usepackage{graphicx}
\usepackage{svg}
\usepackage{subfigure} 
\usepackage{multirow}

\usepackage[font=footnotesize, skip = 1pt, labelfont=bf]{caption}
\usepackage[font=footnotesize]{subcaption}
\begin{document}
\title{Deep Learning in Mild Cognitive Impairment Diagnosis using Eye Movements and Image Content in Visual Memory Tasks}

\titlerunning{Deep Learning in MCI using Eye Movements and Image Content}
%
\author{Tom\'as {Silva Santos Rocha}\inst{1}\textsuperscript{ a} \and
Anastasiia {Mikhailova}\inst{1,2,3}\textsuperscript{ b} \and Moreno~I.~{Coco}\inst{4}\textsuperscript{ c} \and
Jos\'e {Santos-Victor}\inst{1}\textsuperscript{ d}}
\authorrunning{T.~S.~S.~Rocha et al.}
%
\institute{Institute for Systems and Robotics, Laboratory for Robotics and Engineering Systems, Instituto Superior Técnico, Universidade de Lisboa, Lisbon, Portugal \\
    \textsuperscript{a} tomasrocha01@tecnico.ulisboa.pt, \textsuperscript{d} jose.santos-victor@tecnico.ulisboa.pt \\
    \and Department of Psychology, University of Chicago, Chicago, USA \\
    \and Institute for Mind and Biology, University of Chicago, Chicago, USA\\
    \textsuperscript{b} amikhailova@uchicago.edu \\
    \and Department of Psychology, Sapienza Università di Roma, Rome, Italy\\
    \textsuperscript{c} moreno.coco@uniroma1.it}
\maketitle              
%

\begin{abstract}
    The global prevalence of dementia is projected to double by 2050, highlighting the urgent need for scalable diagnostic tools.
    This study utilizes digital cognitive tasks with eye-tracking data correlated with memory processes to distinguish between Healthy Controls (HC) and Mild Cognitive Impairment (MCI), a precursor to dementia.
    A deep learning model based on VTNet was trained using eye-tracking data from 44 participants (24 MCI, 20 HCs) who performed a visual memory task.
    The model utilizes both time series and spatial data derived from eye-tracking.
    It was modified to incorporate scan paths, heat maps, and image content.
    These modifications also enabled testing parameters such as image resolution and task performance, analysing their impact on model performance.
    The best model, utilizing \(700 \times 700px\) resolution heatmaps, achieved 68\% sensitivity and 76\%  specificity.
    Despite operating under more challenging conditions (e.g., smaller dataset size, shorter task duration, or a less standardized task), the model’s performance is comparable to an Alzheimer’s study using similar methods (70\% sensitivity and 73\% specificity).
    These findings contribute to the development of automated diagnostic tools for MCI.\@
    Future work should focus on refining the model and using a standardized long-term visual memory task.

\end{abstract}

\section{Introduction\label{sec:intro}}

The WHO's ``Global Status Report on the Public Health Response to dementia''~\cite{world2021global} highlights dementia as a growing global health issue, affecting over 55 million people, projected to rise to 139 million by 2050, with a disproportionate impact expected in middle-income countries.
This emphasizes the urgent need for diagnostic tools and interventions, particularly as populations age.
Dementia is characterized by cognitive decline that interferes with daily life, including memory loss and impaired thinking.
Alzheimer’s disease is the most common form, though other types include vascular, frontotemporal, and Lewy body dementia~\cite{doi:10.1177/0891988706291081}.
Before developing dementia, patients go through a stage called Mild Cognitive Impairment (MCI), experiencing mild cognitive changes that do not significantly interfere with daily life activities.
Early detection of MCI provides a critical window for interventions that may delay progression to dementia~\cite{petersen2016mild}.
There are two types: amnestic, which impacts memory and has a higher risk of Alzheimer’s, and non-amnestic, which affects attention, language, or visuospatial skills and may indicate risk for other dementias, such as frontotemporal or vascular dementia~\cite{roberts2013classification}.
MCI may also involve single-domain or multiple-domain impairments, with the latter suggesting more significant progression to dementia~\cite{roberts2013classification,gauthier2006mild}.

\subsection{Diagnosing MCI}
MCI is influenced by age, genetics (e.g., APOE \(\varepsilon 4\) allele), lifestyle, and cardiovascular health and is associated with biomarkers such as hippocampal atrophy and amyloid-beta accumulation.
However, diagnosing it promptly remains challenging, as neuroimaging and genetic tests are primarily used for the confirmation of the diagnosis, and lifestyle factors lack diagnostic precision~\cite{livingston2024dementia,https://doi.org/10.1016/j.jalz.2018.02.018}.
The limited availability of specialists restricts the widespread use of standard cognitive tests like MMSE and MoCA.\@
When combined with delayed symptom recognition, this often results in MCI being detected only when cognitive decline is noticeable, limiting early intervention opportunities~\cite{bernstein2022facilitators,chen2024correlates}.

To tackle these challenges, digital assessments on computers, tablets, and smartphones are emerging as practical and cost-effective alternatives~\cite{sabbagh2020rationale}.
When integrated with wearable sensors, cameras, or eye-tracking devices, these tools capture many relevant data without adding time costs, enhancing their diagnostic potential~\cite{li2023synergy}.

Eye-tracking integration in MCI diagnosis has gained attention, with studies suggesting that eye movement patterns could serve as early indicators of cognitive decline~\cite{10.3389/fpsyg.2023.1197567}.

\subsection{Eye-movements: A window to the memories}
Eye movements, essential for visual perception, offer insight into cognitive processing and memory.
Key movements include saccades, rapid gaze shifts between points, and fixations, where the eyes pause to process information.
Saccades allow quick repositioning but temporarily suppress visual intake, while fixations enable detailed analysis, with durations varying by task~\cite{reisberg2013oxford}.
Fixations are particularly useful, as their longer periods allow for easier detection.
Eye movements significantly impact memory mechanisms, aiding encoding and retrieval.
Increased fixations and shifts in gaze while encoding a scene improve memory recall, as eye movements facilitate scene exploration~\cite{DAMIANO2019119}.
During encoding, eye movement patterns differ notably between MCIs and healthy individuals.
MCI patients often have delayed saccadic responses and less accurate saccade targeting, leading to inefficient visual exploration.
Their fixations are longer but fewer, possibly indicating attention and visual processing difficulties.
This reduced fixation frequency and longer saccadic latency contribute to limited scene exploration, impacting their ability to recognize and remember visual details~\cite{10.3389/fpsyg.2023.1197567}.
In the literature, we also see that image content influences eye movements by engaging in a process where our eyes selectively focus on elements perceived as necessary, consequently influencing the memory processes~\cite{doi:10.1152/jn.00145.2019}.

\subsection{Deep Learning and MCI}
To enhance the digital tools deep learning algorithms are being employed, improving diagnostic accuracy and scalability, thus making the timely detection of MCI a possibility~\cite{10.3389/fpsyg.2023.1197567}.

Algorithms based on various methods have been employed, such as Recursive Neural Networks (RNN), Convolutional Neural Networks (CNN), autoencoders, transformers, and others~\cite{SHAH2025103202}.
The transformer based architectures usually perform better than the others, however they require large amounts of data to be properly trained~\cite{SHAH2025103202,madan2024transformer}.
One problem with recent studies is the fact that, typically there is no differentiation between MCI patients and patients with more advanced dementias, such as Alzheimer's, which can bring biases to the results obtained~\cite{SHAH2025103202}.

Building upon recent advancements, this study explores deep learning models that use eye-tracking data collected from MCI patients and Healthy Controls (HCs).
The participants perform a visual long-term active memory task to predict MCI, which have been shown to have better diagnostic accuracy~\cite{TADOKORO2021117529,10.3389/fpsyg.2023.1197567}.
Additionally, this paper investigates incorporating image content, a topic that has not been extensively researched, and memory performance to try to enhance the diagnostic process further.

\section{Method\label{sec: Methodology}}

Initially, we planned to use our collected data; however, the process is ongoing, and the sample size is insufficient.
Instead, we utilized data from~\cite{coco2021semantic}, whose experiment closely aligned with our study.
This study~\cite{coco2021semantic} examined how semantically related images impact memory performance in MCI and HC groups.

\subsection{Participants and Procedure}

\textbf{Selection criteria:} (a) ages between 50 and 90; (b) minimum of three schooling years; (c) normal or corrected-to-normal vision with no history of eye surgery; (d) no history of neurological (except memory disturbances) or psychiatric disorders; (e) no history of alcohol/substance abuse or medications affecting cognition.

Data collection resulted in 44 participants (24 MCI patients: \(71.92 \pm 9.06 \) years old, \(9.83 \pm 4.50\) schooling years; 20 HCs: \(68.50 \pm 8.79 \) years old, \(11.05 \pm 5.10\) schooling years), and a total of 79 sessions after filtering for (a) follow-up participation; (b) performance above chance; (c) successful eye-tracker calibration.

The visual memory task was conducted on a 17-inch monitor (\(1600\times900\) resolution, 50 Hz refresh rate).
Eye movements were recorded with an EyeTribe eye-tracker (55 Hz sampling), with accuracy validated in~\cite{jimaging4080096}, and calibrated at a low resolution (55 Hz), which corresponds to a sample every 33 ms.
The experiment was implemented using OpenSesame~\cite{mathot2012opensesame} and PyGaze~\cite{dalmaijer2014pygaze} to calibrate the eye-tracking using a 9-point procedure before each session.

\begin{figure}[tb]
    \centering
    \includegraphics[width=0.6\textwidth]{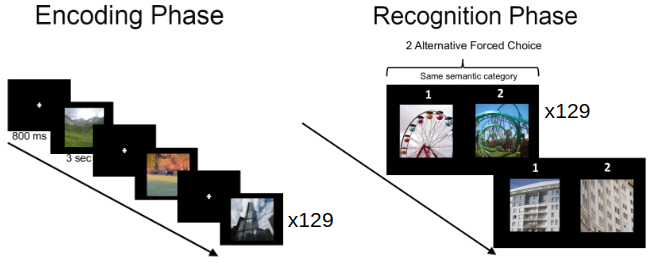}
    \caption{Schematic of the experiment performed by the participants.
        Left: Encoding Phase; Right: Recognition Phase.~\cite{coco2021semantic}\label{fig:experiment}}
\end{figure}

Each session, approximately 30 minutes, consisted of one encoding and recognition phase (Fig.~\ref{fig:experiment}).
During encoding, participants viewed 129 real-life images (\(700\times700\) pixels) randomly selected from 10 lists (834 images total), each shown for 3 seconds with an 800 ms interval between trials.

In the recognition phase, participants saw two images side by side, one of which was previously seen, and were asked to identify the previously seen image.

\subsection{Data Curation and Processing}
Pre-processing addressed missing eye-tracking data from blinks, movements, or detection errors, removing fixations outside the image area~\cite{HESSELS2019100710}.
Trials with \( \geq 20\% \) missing data or gaps over one second were discarded; otherwise, only lost points were removed~\cite{krstic2018all}.
This reduced the dataset from 10,191 (4,773 HCs, 5,418 MCI) to 8,615 trials (4,225 HCs, 4,390 MCI).
To mitigate imbalance, data from the best-performing MCI patient was excluded, further refining the dataset to 8,499 trials (4,225 HCs, 4,274 MCI)~\cite{BHARGAVA20241964,oladunni2021covid}.

The final dataset was divided into 10 balanced folds for a 10-fold nested cross-validation for model evaluation and hyperparameter tuning.
Data from each participant was assigned to a single fold to prevent bias.
Each fold contained a training set (\( \sim80\% \); 35 participants: 19 MCIs and 16 HCs), validation set and test set (10\% each; \( \sim 5\) participants: 3 MCIs and 2 HCs each).

\subsection{The Model}

We selected the VTNet~\cite{10.1145/3382507.3418828} due to its prior success in Alzheimer's prediction~\cite{sriram2023classification}.
Model training and testing were conducted using PyTorch 2.3.0+cu121~\cite{10.1145/3620665.3640366} with the NVIDIA\textsuperscript{\textregistered} CUDA\textsuperscript{\textregistered} Toolkit 12.1~\cite{cuda}, running on an Intel\textsuperscript{\textregistered} Core\texttrademark~i7-9700 CPU, 64 GB RAM, and an NVIDIA\textsuperscript{\textregistered} GeForce RTX\texttrademark~3090 GPU (24 GB memory).
The code is available on GitHub\footnote[1]{\url{https://github.com/tometaro07/eye-tracking-mci.git}}.

\subsubsection{Architecture}

VTNet, Fig.~\ref{fig:architectures} consists of three key components: an RNN path, a CNN path, and a classifier.
The RNN path processes gaze position and pupil size time-series (x, y axes for both eyes) across 107 time steps using a self-attention layer followed by a Gated Recurrent Unit (GRU) to capture sequential patterns.
The CNN path processes a visual representation, with two convolutional layers and max pooling, encodes spatial gaze distributions.
Outputs from both paths are concatenated and passed through two fully connected layers for trial-wise predictions, which can be aggregated at the participant level.

\subsubsection{Proposed Models}

To refine the base model~\cite{sriram2023classification}, we explored several adaptations.
First, scanpath images (\(700\times700px\)), visualizing gaze sequences, were resized to \(256\times256px\) to more closely align with the model’s expected dimensions (\(230\times350 px\)).
We then compared scanpaths with gaze heatmaps, which highlight fixation density using a Multivariate Normal Distribution (Equation~\ref{form:heatmap}):

\begin{equation}
    \label{form:heatmap}
    f(x,y) = \frac{
        \exp\left(
        -\frac{1}{2}
        \left[
            {\left( \frac{x-\mu_x}{\sigma_x} \right)}^2
            + {\left( \frac{y-\mu_y}{\sigma_y} \right)}^2
            \right]
        \right)
    }
    {2\pi \sigma_x \sigma_y}
\end{equation}

Where \( \mu \) is the gaze point and \( \sigma \) is a standard deviation of 1\textdegree{} visual angle.

\begin{figure}[tb]
    \centering
    \begin{tabular}[]{cc}
        \subfigure[]{\includegraphics[width=0.37\textwidth]{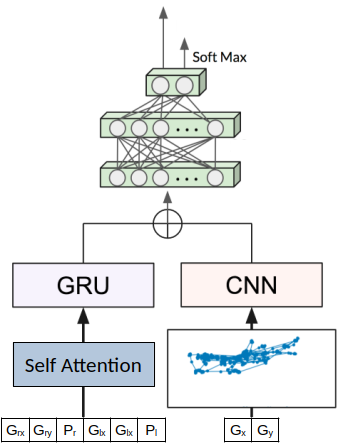}\label{fig:architectures}}
         &
        \begin{tabular}[b]{c}
            \begin{tabular}[b]{cc}
                \subfigure[]{\includegraphics[width=0.14\textwidth]{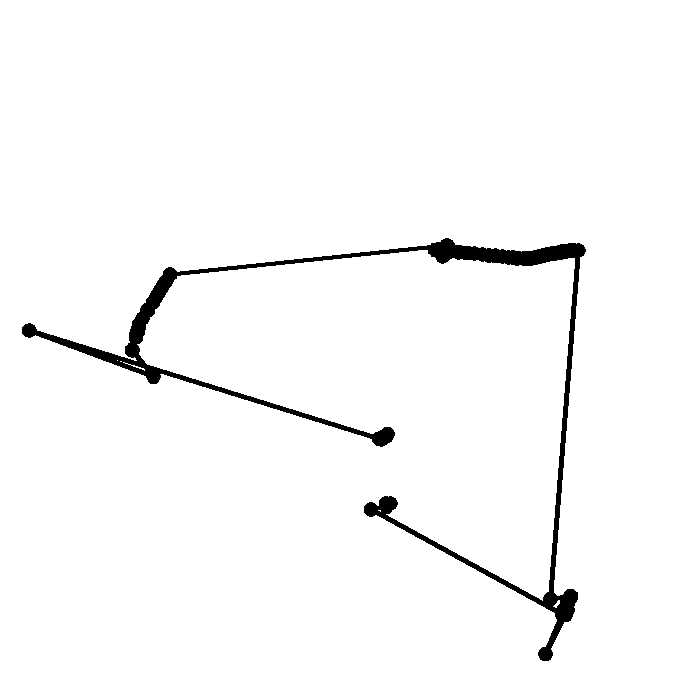}\label{fig:scanpath}}
                 &
                \subfigure[]{\includegraphics[width=0.14\textwidth]{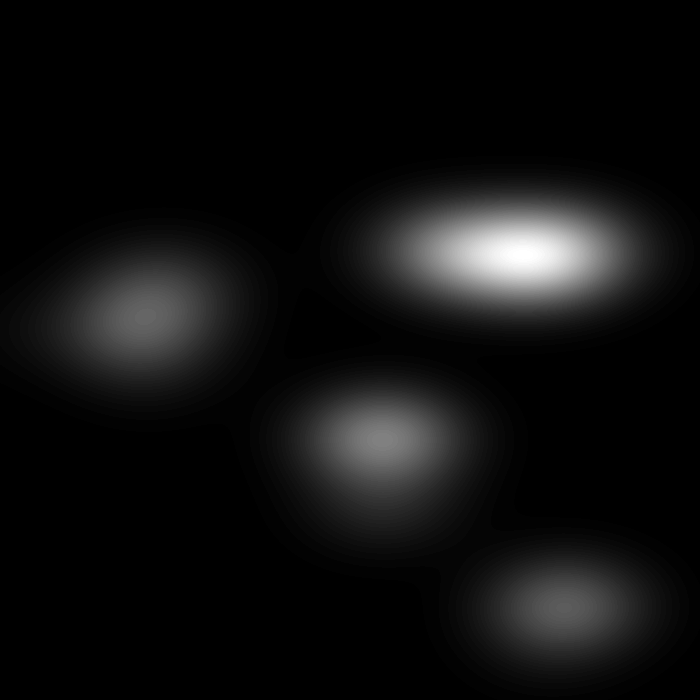}\label{fig:heatmap}}
                \\
                \subfigure[]{\includegraphics[width=0.14\textwidth]{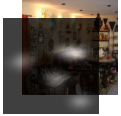}\label{fig:4d}}
                 &
                \subfigure[]{\includegraphics[width=0.14\textwidth]{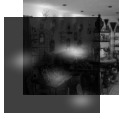}\label{fig:bw-heat}}
            \end{tabular}
            \\
            \subfigure[]{\includegraphics[width=0.14\textwidth]{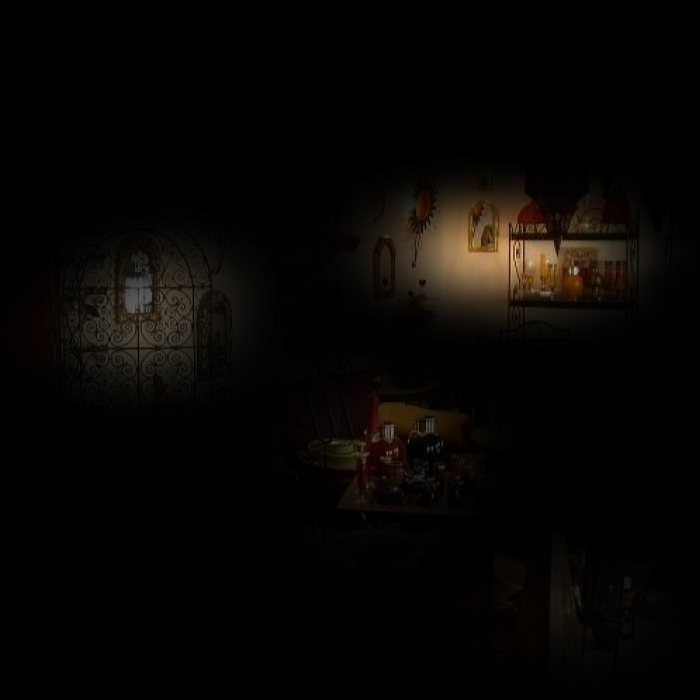}\label{fig:im-mix}}
        \end{tabular}
    \end{tabular}
    \caption{Architecture for the models described and examples of the different types of inputs used for the CNN pathway.
        (a) Architecture: Left path - RNN path, with Self-Attention Layer and GRU Layer (\(G_{rx}\) and \(G_{ry}\) coordinates of the gaze for the right eye; \(G_{lx}\) and \(G_{ly}\) coordinates of the gaze for the left eye; \(P_{x}\) and \(P_{l}\) pupil size for left and right eye); Left Path - CNN Path (\(G_{x}\) and \(G_{y}\) coordinates of gaze); Green Part - Classifier~\cite{10.1145/3382507.3418828}; (b) Scanpath; (c) Heatmap; (d) 4D Image; (e) GS image and Heatmap; (f) 3D Image multiplied by Heatmap mask}%
    \label{fig:inputs}%
\end{figure}

Next, we investigated whether image content could aid MCI prediction by integrating visual representations.
Three approaches were tested: (1) a 4D image combining RGB channels with a heatmap mask (Fig.~\ref{fig:4d}), (2) a 2D Gray Scale, GS, image with a separate heatmap channel (Fig.~\ref{fig:bw-heat}), and (3) an RGB image multiplied by the heatmap mask (Fig.~\ref{fig:im-mix}).

Finally, we examined the impact of task performance on model accuracy using two strategies: one incorporating trial-by-trial performance in the concatenated vector, Single Trial Score (STS) and another combining each participant’s overall score with the model’s average trial performance, Average Test Score (ATS).
These two models combine the task performance with the best-performing models from the ones described above.

\subsection{Training, Architecture Optimization and Model Analysis}
Each model was trained with a batch size 32, balancing computational efficiency and model performance.
We employed the Cross-Entropy Loss function and the Adam optimizer.
The initial learning rate was set to \(10^{-5}\), as higher rates resulted in convergence failure.
The learning rate was halved after five consecutive validation epochs without loss reduction, down to a minimum of \(10^{-6}\) to facilitate fine-tuning.
Early stopping was implemented to prevent overfitting, halting training after 10 epochs without improvement in validation loss and restoring the model to its best-performing weights.

Hyperparameter tuning focused on optimizing the number of convolutional layer channels (6, 8, 12, 16) and the output vector sizes for the RNN and CNN paths (8, 16, 32, 64, 128).
The optimal model configuration was determined by maximizing the combined sensitivity and specificity.
To contextualize our findings, we compared our results to those obtained in the Image Description Task from~\cite{sriram2023classification}.
This study, which also utilized VTNet, involved a similar image exploration task but was exclusively applied to Alzheimer's patients.
To provide additional insight into the visual complexity of the input images, we calculated image entropy using the following formula~\cite{doi:10.1080/713821475}:

\begin{equation}
    H = -\sum_{i=0}^{n-1} p_{i}\log_2 p_{i}
\end{equation}

\section{Results \& discussion\label{sec:resul}}

\subsection{Scanpaths}

We first analysed models similar to the original study~\cite{sriram2023classification}, using scanpath inputs of \(256 \times 256px\) and \(700 \times 700px\).
Results are presented in Table~\ref{tab:scanpath-results}.

The \(256 \times 256px\) model exhibited lower sensitivity (53\%) in differentiating between MCI patients and HCs, compared to the Image Description Task~\cite{sriram2023classification}, the base study (70\%).
This difference is likely due to subtler distinctions between MCI and healthy controls (HCs), compared to the differences between dementia and HC classification in the original study.
Additionally, the higher standard deviation (\(256 \times 256px\) model: \(\sim \)41\%; Base study: 0.02\%) may reflect the smaller sample size (our dataset: 44 participants; Base study dataset: 144 participants), making misclassification more impactful.

Comparing input sizes, the \(700 \times 700px\) model performed worse, likely because scanpath images are 98\% empty space, making smaller images more effective at highlighting relevant patterns and preventing overfitting.
Moreover, the original architecture was designed with \(230 \times 350px\) inputs in mind, reinforcing the importance of choosing an appropriate image size, a trend also observed in handwriting recognition and image segmentation models~\cite{semma2021writer,rukundo2023effects}.

\begin{table*}[tb]
  \centering
  \caption{Mean Sensitivity and specificity, in percentage, with respective standard deviation, for the models using scanpaths and heatmaps as input.
    The results from the image description task in the original study~\cite{sriram2023classification} are also included.}\label{tab:scanpath-results}\label{tab:heatmap-results}
\begin{tabular}{lll|ll|l}
                                     & \multicolumn{2}{l|}{Scanpath}             & \multicolumn{2}{l|}{Heatmap} & \multirow{2}{*}{\begin{tabular}[c]{@{}l@{}}Image Description\\ Task~\cite{sriram2023classification}\end{tabular}}                                           \\ \cline{2-5}
                                     & \multicolumn{1}{l|}{(\(256 \times 256\))} & (\(700 \times 700\))         & \multicolumn{1}{l|}{(\(256 \times 256\))}                                                                         & (\(700 \times 700\)) &                  \\ \hline
    \multicolumn{1}{c|}{Sensitivity} & \multicolumn{1}{l|}{\(52.63 \pm 37.91\)}  & \(47.37 \pm 39.53\)          & \multicolumn{1}{l|}{\(63.16 \pm 28.50\)}                                                                          & \(68.42 \pm 27.26\)  & \(70 \pm 0.02\)  \\
    \multicolumn{1}{c|}{Specificity} & \multicolumn{1}{l|}{\(70.56 \pm 41.08\)}  & \(58.82 \pm 41.50\)          & \multicolumn{1}{l|}{\(52.94 \pm 43.46\)}                                                                          & \(76.47 \pm 27.64\)  & \( 73 \pm 0.02\)
\end{tabular}

\end{table*}

\subsection{Heatmaps}
Contrary to the scanpath models, the \(700 \times 700px\) heatmap model outperformed the \(256 \times 256px\) one due to the loss of information when resizing heatmaps, which led to a performance drop, as shown in Table~\ref{tab:heatmap-results}.
Research indicates that increasing image size benefits performance up to a point, after which gains plateau~\cite{doi:10.1148/ryai.2019190015}.

\begin{figure}[tb]%
  \centering
  \subfigure[Image seen]{\includegraphics[width=0.21\textwidth]{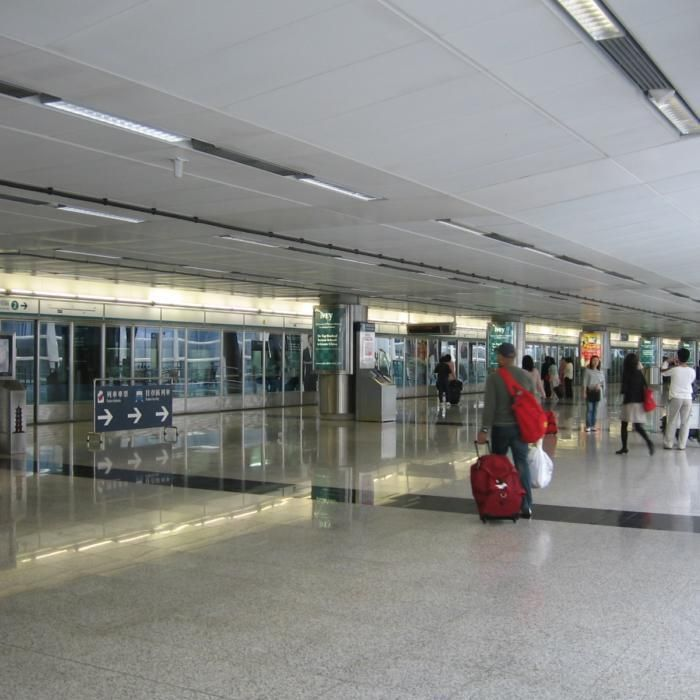}\label{fig:example}}
  \qquad
  \subfigure[Gaze Heatmap of HCs]{\includegraphics[width=0.21\textwidth]{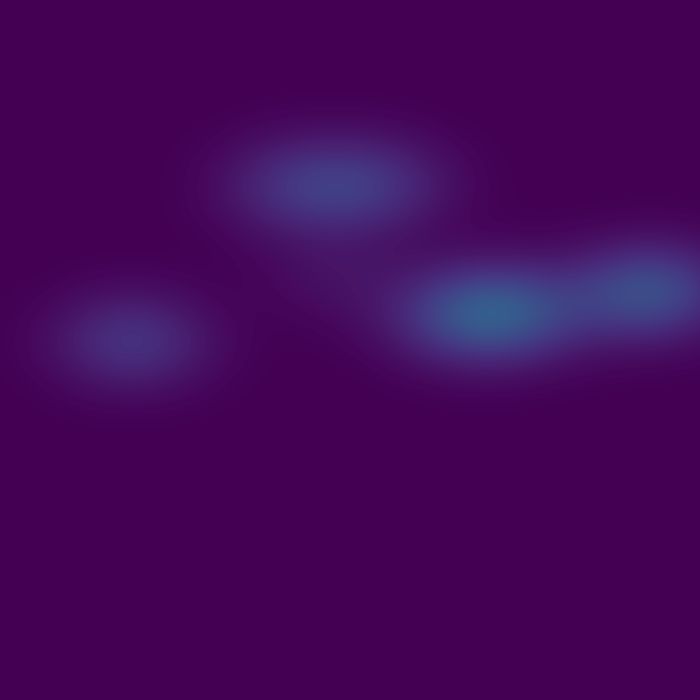}\label{fig:heat-good} }
  \qquad
  \subfigure[Gaze Heatmap of MCI patient]{\includegraphics[width=0.21\textwidth]{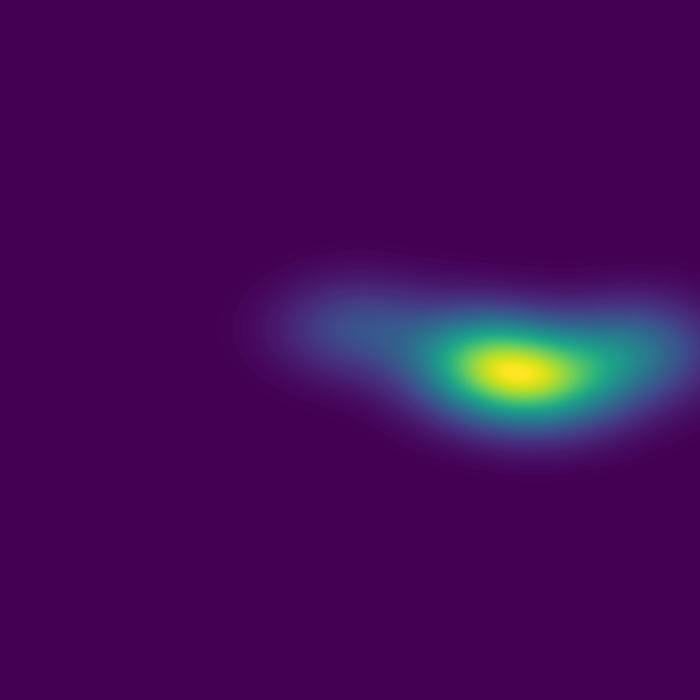}\label{fig:heat-bad} }
  \caption{Examples of the types of inputs used to test if image content can improve the model's result}%
  \label{fig:heatmap-difs}%
\end{figure}

To compare the best models from the heatmap and scanpath models, the \(700 \times 700px\) heatmap model and the \(256 \times 256px\) scanpath model, we first have to examine the image entropy of the model's inputs.
Using Fig.~\ref{fig:heatmap-difs} as an example, we can see the gaze heatmaps from two different participants (one MCI, Fig.~\ref{fig:heat-bad}, and one HC, Fig.~\ref{fig:heat-good}) viewing the same scene, Fig.~\ref{fig:example}.
The entropies of these two heatmaps are 2.65 for the HC and 1.8 for the MCI, indicating a lesser spread of the MCI gaze, and consequently, lower image exploration.

The entropy analysis across trials (summarized in Table~\ref{tab:entropy-heat}) revealed that HCs consistently had higher entropy values than MCI patients, highlighting differences in encoding success.
Additionally, correct trials exhibited higher entropy, especially in HCs, suggesting greater exploration improves memory encoding.

\begin{table}[tb]
  \centering
  \caption{Average entropy values for the \(700 \times 700px\) heatmaps and \(256 \times 256px\) scanpath, with respective standard deviation.
    These values are calculated across the participants' group (HCs and MCI patients), and also by grouping the trials in which participants choose the correct answer or not.}
\begin{tabular}{l|lll|lll}
                                                                                     & \multicolumn{3}{l|}{Heatmap \(700 \times 700px\)}                                    & \multicolumn{3}{l}{Scanpath \(256 \times 256px\)}                                                                                                                                                                                                                                                                                                                                                      \\ \cline{2-7}
                                                                                     & \multicolumn{1}{l|}{HCs}                                                             & \multicolumn{1}{l|}{MCI Patients}                                                    & \textbf{Total}                                                  & \multicolumn{1}{l|}{HCs}                                                             & \multicolumn{1}{l|}{MCI Patients}                                                    & \textbf{Total}                                                  \\ \hline
    \multicolumn{1}{c|}{\begin{tabular}[c]{@{}c@{}}Got Trial\\ Correct\end{tabular}} & \multicolumn{1}{l|}{\begin{tabular}[c]{@{}l@{}}2.614\( \pm \) \\ 0.443\end{tabular}} & \multicolumn{1}{l|}{\begin{tabular}[c]{@{}l@{}}2.512\( \pm \) \\ 0.441\end{tabular}} & \begin{tabular}[c]{@{}l@{}}2.567\( \pm \) \\ 0.445\end{tabular} & \multicolumn{1}{l|}{\begin{tabular}[c]{@{}l@{}}0.368\( \pm \) \\ 0.146\end{tabular}} & \multicolumn{1}{l|}{\begin{tabular}[c]{@{}l@{}}0.377\( \pm \) \\ 0.180\end{tabular}} & \begin{tabular}[c]{@{}l@{}}0.372\( \pm \) \\ 0.163\end{tabular} \\ \hline
    \multicolumn{1}{c|}{\begin{tabular}[c]{@{}c@{}}Got Trial\\ Wrong\end{tabular}}   & \multicolumn{1}{l|}{\begin{tabular}[c]{@{}l@{}}2.494\( \pm \) \\ 0.476\end{tabular}} & \multicolumn{1}{l|}{\begin{tabular}[c]{@{}l@{}}2.415\( \pm \) \\ 0.443\end{tabular}} & \begin{tabular}[c]{@{}l@{}}2.446\( \pm \) \\ 0.458\end{tabular} & \multicolumn{1}{l|}{\begin{tabular}[c]{@{}l@{}}0.350\( \pm \) \\ 0.157\end{tabular}} & \multicolumn{1}{l|}{\begin{tabular}[c]{@{}l@{}}0.347\( \pm \) \\ 0.175\end{tabular}} & \begin{tabular}[c]{@{}l@{}}0.348\( \pm \) \\ 0.169\end{tabular} \\ \hline
    \textbf{Total}                                                                   & \multicolumn{1}{l|}{\begin{tabular}[c]{@{}l@{}}2.589\( \pm \) \\ 0.453\end{tabular}} & \multicolumn{1}{l|}{\begin{tabular}[c]{@{}l@{}}2.480\( \pm \) \\ 0.444\end{tabular}} & \begin{tabular}[c]{@{}l@{}}2.534\( \pm \) \\ 0.454\end{tabular} & \multicolumn{1}{l|}{\begin{tabular}[c]{@{}l@{}}0.364\( \pm \) \\ 0.149\end{tabular}} & \multicolumn{1}{l|}{\begin{tabular}[c]{@{}l@{}}0.367\( \pm \) \\ 0.179\end{tabular}} & \begin{tabular}[c]{@{}l@{}}0.366\( \pm \) \\ 0.165\end{tabular}
\end{tabular}\label{tab:entropy-heat}\label{tab:entropy-scan}
\end{table}

Scanpath entropies (Table~\ref{tab:entropy-scan}) showed minimal differences between classes (0.003), significantly lower than the difference between classes in heatmap entropies (0.109).
This suggests that entropy plays a role in model performance, where higher entropy differences between classes correlate to better results, evidence also supported by~\cite{rahane2020measures} and~\cite{frank2020salient}.

In conclusion, the \(700 \times 700px\) heatmap model demonstrated better performance and higher consistency compared to the \(256 \times 256px\) scanpath model.
When compared to the base study~\cite{sriram2023classification}, the model performance was similar (\(700 \times 700px\) heatmap: Sensitivity-68\%, Specificity-76\%; Base model: Sensitivity-70\%, Specificity-73\%).

\subsection{Image Content}

In the three image content models, Table~\ref{tab:im-content-results}: Heatmap + RGB Image, Heatmap + GS Image, and RGB Image \(\times \) Heatmap, the performance declined, despite the performance of the model using only the \(700 \times 700px\) heatmap.
This decline may be due to a suboptimal architecture for these inputs or overfitting.

\begin{table*}[tb]
  \centering
  \caption{Mean Sensitivity and specificity, in percentage, with respective standard deviation, for the models using image content as input, after the 10-fold nested cross-validation.
    The results from the \(700 \times 700px\) Heatmap model are also included, for comparison.}\label{tab:im-content-results}
  \renewcommand{\arraystretch}{0.8}
\begin{tabular}{c|l|l|l|l}
  \multicolumn{1}{l|}{}                                                             &
  \begin{tabular}[c]{@{}l@{}}Heatmap\\ (\(700 \times 700\))\end{tabular}            &
  \begin{tabular}[c]{@{}l@{}}Heatmap + RGB\\Image (\(700 \times 700\))\end{tabular} &
  \begin{tabular}[c]{@{}l@{}}Heatmap + GS\\Image (\(700 \times 700\))\end{tabular}  &
  \begin{tabular}[c]{@{}l@{}}RGB Image  \(\times \) \\Heatmap (\(700 \times 700\))\end{tabular} \\ \hline
  Sensitivity                                                                       &
  \(68.42 \pm 27.26\)                                                               &
  \(46.34 \pm 47.80\)                                                               &
  \(48.78 \pm 45.64\)                                                               &
  \(63.41 \pm 33.78\)                                                                           \\
  Specificity                                                                       &
  \(76.47 \pm 27.64\)                                                               &
  \(62.16 \pm 44.31\)                                                               &
  \(56.76 \pm 44.87\)                                                               &
  \(43.24 \pm 40.11\)
\end{tabular}
\end{table*}

This result may have also been partially affected by the random image selection, resulting in an uneven distribution, with only 365 images shown more than 10 times and over 100 shown just twice (Fig.~\ref{fig:img-diistri}).
This variability may have confused the model, though generalized eye movement patterns could still be beneficial, albeit requiring a larger dataset.
As such, we suggest that conducting experiments with a more controlled image set could improve MCI prediction outcomes.

\begin{figure}[tb]
  \begin{minipage}[c]{.45\linewidth}
    \centering
    \includesvg[width=\textwidth]{./Images/Section3/image_distribution.svg}
    \caption{Histogram of the distribution of the images used during the experiment.}\label{fig:img-diistri}
  \end{minipage}\hfill
  \begin{minipage}[c]{.43\linewidth}
    \centering
    \includesvg[width=\textwidth]{./Images/Section3/test_score.svg}
    \caption{Boxplot of the Test score for each of the experiment runs, separated by groups.}\label{fig:test_score}
  \end{minipage}
\end{figure}

\subsection{Trial and Test Score}

We then analysed the two models related to task performance, Table~\ref{tab:score-results}: Heatmap + STS and Heatmap + ATS.\@
Both models underperformed compared to the \(700 \times 700px\) gaze heatmap model, though the trial score model’s sensitivity was similar to the heatmap model, with reduced specificity.
This suggests the model over-classifies HCs as MCI patients, possibly due to misclassifications in trial recognition.
The entropy analysis of heatmaps (Table~\ref{tab:entropy-heat}) also supports that incorrect trials exhibit smaller group differences, making trial scores less beneficial.

\begin{table*}[tb]
  \centering
  \caption{Mean Sensitivity and specificity, in percentage, with respective standard deviation, for the models using the scores from both each trial and the average of the whole test as input, after the 10-fold nested cross-validation.
    The results from the \(700 \times 700px\) Heatmap model are also included, for comparison.}\label{tab:score-results}
  \renewcommand{\arraystretch}{0.8}
\begin{tabular}[b]{c|l|l|l}
  \multicolumn{1}{l|}{}                                                            &
  \begin{tabular}[c|]{@{}l@{}}Heatmap\\ (\(700 \times 700px\))\end{tabular}        &
  \begin{tabular}[c|]{@{}l@{}}Heatmap + STS \\ (\(700 \times 700px\))\end{tabular} &
  \begin{tabular}[c|]{@{}l@{}}Heatmap + ATS\\ (\(700 \times 700px\))\end{tabular}    \\ \hline
  Sensitivity                                                                      &
  \(68.42 \pm 27.26\)                                                              &
  \(68.29 \pm 42.57\)                                                              &
  \(56.10 \pm 49.72\)                                                                \\
  Specificity                                                                      &
  \(76.47 \pm 27.64\)                                                              &
  \(45.95 \pm 40.10\)                                                              &
  \(40.54 \pm 51.64\)
\end{tabular}
\end{table*}

For the test score model, despite marked differences in average test scores between HCs (78.8\%) and MCI patients (66.8\%), there was significant overlap in performance, with MCI patients showing more variability (Fig.~\ref{fig:test_score}).
Two factors limiting performance were the overlap in test scores and the small sample size of around 35 participants per training set, which constrained the model's learning capability.

\section{Conclusions}\label{sec:concl}

This paper explored the use of deep learning to predict MCI using eye-tracking data and image content in a visual memory task.
The \(700 \times 700px\) heatmap model proved to be the most effective, achieving 68\% sensitivity and 76\% specificity.
These results are comparable to those of the base study~\cite{sriram2023classification} (70\% Sensitivity and 73\% specificity), despite our study operating under more challenging conditions.
However, the performance remains below diagnostic thresholds.

The models using image content performed significantly lower than expected.
We believe this was primarily due to the large number of images relative to the sample size, which hindered the models' ability to extract meaningful information.

Furthermore, incorporating trial-by-trial and overall task scores did not improve the results.
This suggests that the gaze heatmaps may already encapsulate relevant information regarding trial-wise performance, rendering the direct use of task scores redundant.

Future work should prioritize standardizing the image set across all participants and increasing the sample size.
Additionally, adapting the model for use with digital cameras could enhance accessibility for MCI screening.

\begin{credits}\label{sec:credits}
    \subsubsection{\ackname}
    This study is funded by Lisbon ELLIS Unit, the Center for Responsible AI (PRR), LARSyS FCT funding (DOI: 10.54499/LA/P/0083/2020, \allowbreak{10.54499/UIDP/50009/2020}, and 10.54499/UIDB/50009/2020), and was supported by grants from NVIDIA and utilized NVIDIA\textsuperscript{\textregistered} GeForce RTX\texttrademark~3090 GPU.
    \subsubsection{\discintname}
    All authors have no conflicts of interest.
\end{credits}

%
%


\begin{thebibliography}{10}
\providecommand{\url}[1]{\texttt{#1}}
\providecommand{\urlprefix}{URL }
\providecommand{\doi}[1]{https://doi.org/#1}

\bibitem{10.1145/3620665.3640366}
Ansel, J., Yang, E., He, H., Gimelshein, N., Jain, A., Voznesensky, M., Bao, B., Bell, P., Berard, D., Burovski, E., Chauhan, G., Chourdia, A., Constable, W., Desmaison, A., DeVito, Z., Ellison, E., Feng, W., Gong, J., Gschwind, M., Hirsh, B., Huang, S., Kalambarkar, K., Kirsch, L., Lazos, M., Lezcano, M., Liang, Y., Liang, J., Lu, Y., Luk, C.K., Maher, B., Pan, Y., Puhrsch, C., Reso, M., Saroufim, M., Siraichi, M.Y., Suk, H., Zhang, S., Suo, M., Tillet, P., Zhao, X., Wang, E., Zhou, K., Zou, R., Wang, X., Mathews, A., Wen, W., Chanan, G., Wu, P., Chintala, S.: Pytorch 2: Faster machine learning through dynamic python bytecode transformation and graph compilation. In: Proceedings of the 29th ACM International Conference on Architectural Support for Programming Languages and Operating Systems, Volume 2. p. 929–947. Association for Computing Machinery (2024). \doi{10.1145/3620665.3640366}

\bibitem{bernstein2022facilitators}
Bernstein~Sideman, A., Al-Rousan, T., Tsoy, E., Pi{\~n}a~Escudero, S.D., Pintado-Caipa, M., Kanjanapong, S., Mbakile-Mahlanza, L., Okada~de Oliveira, M., De~la Cruz-Puebla, M., Zygouris, S., et~al.: Facilitators and barriers to dementia assessment and diagnosis: Perspectives from dementia experts within a global health context. Frontiers in neurology  \textbf{13},  769360 (2022). \doi{10.3389/fneur.2022.769360}

\bibitem{BHARGAVA20241964}
Bhargava, Y., Shetty, S.K., Baths, V.: Subjective cognitive decline prediction on imbalanced data using data-resampling and cost-sensitive training methods. Procedia Computer Science  \textbf{235},  1964--1979 (2024). \doi{10.1016/j.procs.2024.04.186}, international Conference on Machine Learning and Data Engineering (ICMLDE 2023)

\bibitem{doi:10.1177/0891988706291081}
Breitner, J.C.S.: Dementia—epidemiological considerations, nomenclature, and a tacit consensus definition. Journal of Geriatric Psychiatry and Neurology  \textbf{19}(3),  129--136 (2006). \doi{10.1177/0891988706291081}, pMID: 16880354

\bibitem{chen2024correlates}
Chen, Y., Power, M.C., Grodstein, F., Capuano, A.W., Lange-Maia, B.S., Moghtaderi, A., Stapp, E.K., Bhattacharyya, J., Shah, R.C., Barnes, L.L., et~al.: Correlates of missed or late versus timely diagnosis of dementia in healthcare settings. Alzheimer's \& Dementia  \textbf{20}(8),  5551--5560 (2024). \doi{10.1002/alz.14067}

\bibitem{coco2021semantic}
Coco, M.I., Merendino, G., Zappal{\`a}, G., Della~Sala, S.: Semantic interference mechanisms on long-term visual memory and their eye-movement signatures in mild cognitive impairment. Neuropsychology  \textbf{35}(5), ~498 (2021). \doi{10.1037/neu0000734}

\bibitem{dalmaijer2014pygaze}
Dalmaijer, E.S., Math{\^o}t, S., Van~der Stigchel, S.: Pygaze: An open-source, cross-platform toolbox for minimal-effort programming of eyetracking experiments. Behavior research methods  \textbf{46},  913--921 (2014). \doi{10.3758/s13428-013-0422-2}

\bibitem{DAMIANO2019119}
Damiano, C., Walther, D.B.: Distinct roles of eye movements during memory encoding and retrieval. Cognition  \textbf{184},  119--129 (2019). \doi{10.1016/j.cognition.2018.12.014}

\bibitem{frank2020salient}
Frank, S.J., Frank, A.M.: Salient slices: Improved neural network training and performance with image entropy. Neural Computation  \textbf{32}(6),  1222--1237 (2020). \doi{10.48550/arXiv.1907.12436}

\bibitem{gauthier2006mild}
Gauthier, S., Reisberg, B., Zaudig, M., Petersen, R.C., Ritchie, K., Broich, K., Belleville, S., Brodaty, H., Bennett, D., Chertkow, H., et~al.: Mild cognitive impairment. The lancet  \textbf{367}(9518),  1262--1270 (2006). \doi{10.1016/S0140-6736(06)68542-5}

\bibitem{HESSELS2019100710}
Hessels, R.S., Hooge, I.T.: Eye tracking in developmental cognitive neuroscience – the good, the bad and the ugly. Developmental Cognitive Neuroscience  \textbf{40},  100710 (2019). \doi{10.1016/j.dcn.2019.100710}

\bibitem{https://doi.org/10.1016/j.jalz.2018.02.018}
Jack~Jr., C.R., Bennett, D.A., Blennow, K., Carrillo, M.C., Dunn, B., Haeberlein, S.B., Holtzman, D.M., Jagust, W., Jessen, F., Karlawish, J., Liu, E., Molinuevo, J.L., Montine, T., Phelps, C., Rankin, K.P., Rowe, C.C., Scheltens, P., Siemers, E., Snyder, H.M., Sperling, R., Contributors, Elliott, C., Masliah, E., Ryan, L., Silverberg, N.: Nia-aa research framework: Toward a biological definition of alzheimer's disease. Alzheimer's \& Dementia  \textbf{14}(4),  535--562 (2018). \doi{10.1016/j.jalz.2018.02.018}

\bibitem{krstic2018all}
Krsti{\'c}, K., {\v{S}}o{\v{s}}ki{\'c}, A., Kovi{\'c}, V., Holmqvist, K.: All good readers are the same, but every low-skilled reader is different: an eye-tracking study using pisa data. European Journal of Psychology of Education  \textbf{33},  521--541 (2018). \doi{10.1007/s10212-018-0382-0}

\bibitem{li2023synergy}
Li, A., Li, J., Zhang, D., Wu, W., Zhao, J., Qiang, Y.: Synergy through integration of digital cognitive tests and wearable devices for mild cognitive impairment screening. Frontiers in Human Neuroscience  \textbf{17},  1183457 (2023). \doi{10.3389/fnhum.2023.1183457}

\bibitem{livingston2024dementia}
Livingston, G., Huntley, J., Liu, K.Y., Costafreda, S.G., Selb{\ae}k, G., Alladi, S., Ames, D., Banerjee, S., Burns, A., Brayne, C., et~al.: Dementia prevention, intervention, and care: 2024 report of the lancet standing commission. The Lancet  \textbf{404}(10452),  572--628 (2024). \doi{10.1016/S0140-6736(24)01296-0}

\bibitem{madan2024transformer}
Madan, S., Lentzen, M., Brandt, J., Rueckert, D., Hofmann-Apitius, M., Fr{\"o}hlich, H.: Transformer models in biomedicine. BMC Medical Informatics and Decision Making  \textbf{24}(1), ~214 (2024). \doi{10.1186/s12911-024-02600-5}

\bibitem{mathot2012opensesame}
Math{\^o}t, S., Schreij, D., Theeuwes, J.: Opensesame: An open-source, graphical experiment builder for the social sciences. Behavior research methods  \textbf{44},  314--324 (2012). \doi{10.3758/s13428-011-0168-7}

\bibitem{cuda}
NVIDIA, Vingelmann, P., Fitzek, F.H.: Cuda, release: 12.1 (2024), \url{https://developer.nvidia.com/cuda-toolkit}

\bibitem{oladunni2021covid}
Oladunni, T., Tossou, S., Haile, Y., Kidane, A.: Covid-19 county level severity classification with imbalanced class: A nearmiss under-sampling approach. medRxiv pp. 2021--05 (2021). \doi{10.1101/2021.05.21.21257603}

\bibitem{jimaging4080096}
Ooms, K., Krassanakis, V.: Measuring the spatial noise of a low-cost eye tracker to enhance fixation detection. Journal of Imaging  \textbf{4}(8) (2018). \doi{10.3390/jimaging4080096}

\bibitem{world2021global}
Organization, W.H., et~al.: Global status report on the public health response to dementia (2021), \url{https://digitalcommons.fiu.edu/srhreports/health/health/65/}

\bibitem{petersen2016mild}
Petersen, R.C.: Mild cognitive impairment. CONTINUUM: lifelong Learning in Neurology  \textbf{22}(2),  404--418 (2016). \doi{10.1212/CON.0000000000000313}

\bibitem{rahane2020measures}
Rahane, A.A., Subramanian, A.: Measures of complexity for large scale image datasets. In: 2020 International Conference on Artificial Intelligence in Information and Communication (ICAIIC). pp. 282--287. IEEE (2020). \doi{10.1109/ICAIIC48513.2020.9065274}

\bibitem{reisberg2013oxford}
Reisberg, D.: The Oxford handbook of cognitive psychology, chap. 5. Eye Movements, pp. 69--82. OUP USA (2013). \doi{10.1093/oxfordhb/9780195376746.001.0001}

\bibitem{roberts2013classification}
Roberts, R., Knopman, D.S.: Classification and epidemiology of mci. Clinics in geriatric medicine  \textbf{29}(4),  753--772 (2013). \doi{10.1016/j.cger.2013.07.003}

\bibitem{rukundo2023effects}
Rukundo, O.: Effects of image size on deep learning. Electronics  \textbf{12}(4), ~985 (2023). \doi{10.3390/electronics12040985}

\bibitem{sabbagh2020rationale}
Sabbagh, M.N., Boada, M., Borson, S., Chilukuri, M., Doraiswamy, P., Dubois, B., Ingram, J., Iwata, A., Porsteinsson, A., Possin, K., et~al.: Rationale for early diagnosis of mild cognitive impairment (mci) supported by emerging digital technologies. The journal of prevention of Alzheimer's disease  \textbf{7},  158--164 (2020). \doi{10.14283/jpad.2020.19}

\bibitem{doi:10.1148/ryai.2019190015}
Sabottke, C.F., Spieler, B.M.: The effect of image resolution on deep learning in radiography. Radiology: Artificial Intelligence  \textbf{2}(1),  e190015 (2020). \doi{10.1148/ryai.2019190015}, pMID: 33937810

\bibitem{semma2021writer}
Semma, A., Lazrak, S., Hannad, Y., Boukhani, M., El~Kettani, Y.: Writer identification: The effect of image resizing on cnn performance. The International Archives of the Photogrammetry, Remote Sensing and Spatial Information Sciences  \textbf{46},  501--507 (2021). \doi{10.5194/isprs-archives-XLVI-4-W5-2021-501-2021}

\bibitem{SHAH2025103202}
Shah, H.A., Khalil, S., Andberg, S., Koivisto, A.M., Bednarik, R.: Eye tracking based detection of mild cognitive impairment: A review. Information Fusion  \textbf{122},  103202 (2025). \doi{j.inffus.2025.103202}

\bibitem{10.1145/3382507.3418828}
Sims, S.D., Conati, C.: A neural architecture for detecting user confusion in eye-tracking data. In: Proceedings of the 2020 International Conference on Multimodal Interaction. p. 15–23. ICMI '20, Association for Computing Machinery, New York, NY, USA (2020). \doi{10.1145/3382507.3418828}

\bibitem{doi:10.1152/jn.00145.2019}
Souto, D., Kerzel, D.: Visual selective attention and the control of tracking eye movements: a critical review. Journal of Neurophysiology  \textbf{125}(5),  1552--1576 (2021). \doi{10.1152/jn.00145.2019}, pMID: 33730516

\bibitem{sriram2023classification}
Sriram, H., Conati, C., Field, T.: Classification of alzheimer's disease with deep learning on eye-tracking data. In: Proceedings of the 25th International Conference on Multimodal Interaction. pp. 104--113 (2023). \doi{10.1145/3577190.3614149}

\bibitem{TADOKORO2021117529}
Tadokoro, K., Yamashita, T., Fukui, Y., Nomura, E., Ohta, Y., Ueno, S., Nishina, S., Tsunoda, K., Wakutani, Y., Takao, Y., Miyoshi, T., Higashi, Y., Osakada, Y., Sasaki, R., Matsumoto, N., Kawahara, Y., Omote, Y., Takemoto, M., Hishikawa, N., Morihara, R., Abe, K.: Early detection of cognitive decline in mild cognitive impairment and alzheimer's disease with a novel eye tracking test. Journal of the Neurological Sciences  \textbf{427},  117529 (2021). \doi{10.1016/j.jns.2021.117529}

\bibitem{doi:10.1080/713821475}
Thum, C.: Measurement of the entropy of an image with application to image focusing. Optica Acta: International Journal of Optics  \textbf{31}(2),  203--211 (1984). \doi{10.1080/713821475}

\bibitem{10.3389/fpsyg.2023.1197567}
Wolf, A., Tripanpitak, K., Umeda, S., Otake-Matsuura, M.: Eye-tracking paradigms for the assessment of mild cognitive impairment: a systematic review. Frontiers in Psychology  \textbf{14} (2023). \doi{10.3389/fpsyg.2023.1197567}

\end{thebibliography}
\end{document}